%% file: localization.tex
\documentclass[11pt]{article}
\newcommand{\doublespace}{\addtolength{\baselineskip}{.25\baselineskip}}
\usepackage{graphicx}
\usepackage{epsfig}

\newtheorem{theorem}{Theorem}

\newtheorem{definition}[theorem]{Definition}

\begin{document}
\title{
{\small\textsf{University of Iowa Department of Computer Science Technical Report TR-01-06}} 
\\
Localization in Wireless Sensor Grids\thanks{The research
presented in this paper is supported 
by the DARPA contract F33615-01-C-1901}} 
\author{Chen Zhang, Ted Herman
\\Department of Computer Science
\\University of Iowa, Iowa City, IA 52242
\\\{chen-zhang, ted-herman \}@uiowa.edu}

\maketitle
\begin{abstract}
This work reports experiences on using radio ranging 
to position sensors in a grid topology.  The implementation is simple, 
efficient, and could be practically distributed.  The 
paper describes an implementation and experimental result based on
RSSI distance estimation. Novel techniques such as fuzzy membership
functions and table lookup are used to obtain more accurate result
and simplify the computation. An 86\% accuracy is achieved in the
experiment in spite of inaccurate RSSI distance estimates with 
errors up to 60\%.    
\end{abstract}

\doublespace

\input{sec1}
\input{sec2}

\input{sec3}
\input{sec4}
\input{sec5}
\input{sec6}

\bibliography{localization}
\bibliographystyle{plain}
\end{document}

%% file: sec1.tex
\section{Introduction}
Sensor networks fundamentally depend on location information for
applications, which must bind spatial coordinates with sensed 
data.  In addition, location can be useful for routing, 
power management, and other services. 
A localization (or positioning) service enables sensor nodes to 
acquire their spatial coordinates without having to program and deploy
each sensor to a precise location.  

In some cases position can be obtained by having Global Positioning
System (GPS) on each sensor node, however this can have high expense, 
power cost, and put problematic constraints on deployment.  
Instead of using GPS, most self-localization algorithms assume
only some sensor nodes can obtain their position information directly, 
and they are called anchors or beacons (by contrast, some algorithms
construct relative coordinates only, and may use constraint solving
techniques to combine local maps \cite{MLRT04}).
Using anchor positions, positions of other sensor 
nodes will then be calculated.  Already there is extensive literature
on the general problem of localization in wireless sensor networks, 
with solutions ranging from specialized ranging assumptions 
to complex algorithms that minimize overall positioning error 
\cite{LR03,NN01,SLR02,SPS02,KMSKA04,BY04,SKMKA05,ZC04,BP00,MSD05,GKLP04,HHBSA03,PS05}.    

The general problem of (self-) localization makes few assumptions about the 
topology of the network beyond basic communication connectivity and
density (both sensor and anchor) constraints:  the topology can be 
arbitrary and sensor deployment could be some random 
process.  Presently, sensors are often expensive enough to justify more 
careful deployment than a random one;  an economical emplacement of 
sensors for uniform coverage is a regular tiling \cite{DHL04} 
or grid topology. In \cite{BKA05}, a large-scale wireless sensor
grid is implemented and evaluated. Some localization schemes \cite{SS04},
\cite{SKMKA05} are also based on the grid deployment.
The question driving our investigation is:  
does the assumption of a grid topology reduce
the localization problem to a simple, practically implementable one?
In a wireless sensor grid, one has more position constraint
information than in a randomly placed network.  With this 
additional information, we hope for a fast localization algorithm 
with small memory footprint, that could reasonably be run on the 
current generation of sensor nodes. 

\textit{Contribution.} 
In this paper, we investigate a very simple approach to 
localization for wireless sensor networks in a grid topology.  
The algorithm consumes little computing power and
does not depend on high precision in distance estimation. It
can be implemented in a distributed manner at different levels. 
Despite its almost na\"{i}ve approach to assigning grid positions
based on table lookup from hop distance and despite significant radio
ranging errors, we find reasonable results 
(the reader may look ahead to Figure \ref{result} for 
illustrative results).  Our experimental results were obtained
in rough and grassy terrain using RSSI for ranging, 
where acoustic ranging --- the foundation for other 
distributed localization research \cite{MLRT04} --- 
has been difficult to use \cite{KMSKA04}.

The rest of the paper is organized as follows. Section~\ref{prelim} describes the 
grid topology and anchor placement. Section~\ref{sec-algorithm} introduces
a table lookup algorithm for localization and discusses different
implementation possibilities.
Section~\ref{sec-experiment} presents experimental results and compares our  
algorithm to related results.
Section~\ref{sec-analysis} provides analysis of
the algorithm, and Section \ref{conclusion} contains  conclusions and future work.

%% file: sec2.tex
\section{Preliminaries}
\label{prelim}
\subsection{Grid Topology and Anchor Placement}
We assume a grid topology with a rectangular perimeter;  
the grid can be divided into segments such that each segment has 
its own set of anchors.
In our initial implementation, each segment has a basestation 
to run the localization algorithm.  We discuss later how this 
assumption could be removed to get a fully distributed implementation
with no basestation.  The output of localization 
is an assignment of sensor nodes to the grid positions. 
 
The size of each segment will affect the number of anchor nodes in the
whole network and the performance of the algorithm.  
When each segment contains fewer nodes, and still has same number
of anchors, then
more anchors are needed for the whole grid. But the localization 
will be faster in this case
and have better accuracy.
 
\begin{figure}[ht]
\centering
\includegraphics {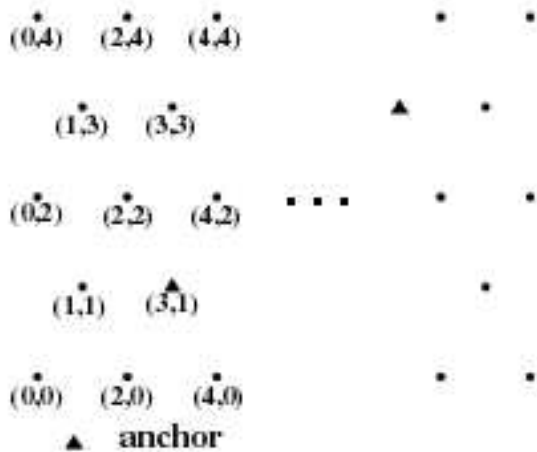}
\caption{segment topology}
\label{segment}
\end{figure}

We field-tested our algorithm in a 5 by 10 grid segment. 
Figure~\ref{segment} shows the topology, where the $(x,y)$ 
position of some points is shown.  This topology
is similar to the segment topology suggested in \cite{BKA05},
which put vertices of the grid in a roughly hexagonal pattern.
In this layout, the unit length of $y$-axis is twice the unit length 
of $x$-axis.  This guarantees that each grid edge (that is, the 
line between nearest points) is of length at most 2:  as 
explained in \cite{BKA05}, this design gives each sensor sufficiently
many communication neighbors to achieve desired redundancy for 
multi-hop routing, while efficiently providing sensor coverage of
the entire field.  

In our test, the segment has four anchors in the segment 
at grid positions (12,0), (3,1), (17,3), and (8,4).  
Anchors need to be placed carefully to ensure sufficient accuracy in 
localization.  We provide some constraints on anchor placement in
section~\ref{sec-algorithm}, however the general problems of optimizing
the number of anchors and optimizing anchor placement remain open.

The error in deployment for our experiment 
is discussed in section~\ref{sec-experiment}.

\subsection{Grid Distance}
In our algorithm, we will use the number of grid hops 
on the shortest path between two nodes to represent the distance
between them. The advantage of this representation is that it reduces
the complexity of computing in localization (for instance, no 
floating-point calculation is needed). 
For two grid positions, it is also easy to find the formula calculating 
distance between them. For two positions (a,b) and (c,d), the distance
between them is given by 
\begin{eqnarray} \label{eqn-1} 
\textit{dist}[\,(a,b),\;(c,d)\,] \quad  =  \quad 
 \max\left(\frac{(|a-c|+|b-d)}{2},|b-d|\right)
\end{eqnarray}

%% file: sec3.tex
\section{Localization Algorithm and Implementation} 
\label{sec-algorithm}
Our localization algorithm is a simple algorithm using 
table lookup.  The entire procedure has three stages. 
The first stage is to calculate
a table, where each grid position has an entry. The entry for a
grid position $p$ is a 4-tuple, consisting of grid distances between
$p$ and four anchors.  At least three anchors are necessary for 
localization;  we found four anchors to be an improvement over 
three anchors, and the scheme can easily be generalized to a 
larger number of anchors.  The second stage establishes the 
one-hop neighborhood for each sensor node, and for each 
node $q$, calculates a 4-tuple, consisting of grid distances 
between $q$ and the anchors using the
one-hop neighborhood information. The third stage performs the table lookup 
and then refines the result of the lookup.
Our algorithm is shown in Figure ~\ref{top-level}.

\begin{figure}[ht]
\centering
\begin{tabbing}
xxx \= xxxxxxxx \= xxx \= xxx \= \kill
\>\bf\em{stage 1: }\\
\>

calculate grid distance table for the grid \\ \>\>positions\\
\\
\>\bf\em{stage 2:}\\
\>
each node $p$ sends a group of 30 messages\\
\>
each node forwards all the RSSI readings to\\
\>\> the base station \\
\>
for each sender $p$, base station calculates \\
\>\>a score for all receivers \\
\>
decide one-hop neighborhood for all nodes \\
\\

\>\bf\em{stage3:}\\
\>calculate distance between each node $q$ \\
\>\>and the anchors\\
\>for each node $q$,look up in the table for\\
\>\> a match  \\
\>for each unoccupied position, find the \\
\>\> most likely node\\
\>send out the assignments

\end{tabbing}
\caption{localization algorithm}
\label{top-level}
\end{figure}

The implementation of this algorithm in one segment can be
either centralized or distributed. We will describe the centralized 
implementation first. Also each stage of this algorithm can be implemented
differently. For example, in the second stage, different ranging techniques
can be used to help establishing the one-hop neighborhood.

\subsection{Table Establishment and Anchors}
Using formula (\ref{eqn-1}), defined in Section \ref{prelim}, 
it is simple to calculate the distances from any grid position $p$ 
to the anchors;  therefore when the grid is designed or deployed the 
table can be calculated and disseminated as needed for the 
subsequent stages.

In order for table lookup to be unambiguous, no pair of tuples
in the table should be identical (otherwise two different grid points could
have the same lookup). This constrains anchor placement.
If anchors are placed in a line, then there will be much symmetry in the 
segment and different positions have identical tuples, which should
be avoided. In our experiment, we set up the segment same as in \cite{BKA05},
 we found that if the 4 anchors
form a parallelogram, then no table entries are identical. We have 
simulated different parallelogram placements in one segment, and
found when the anchors are placed close to the segment border, the 
performance is slightly better. Similar observations can be found in 
other localization schemes \cite{SGAMS03}.

\subsection{One-hop Neighborhood using RSSI}

We use Received Signal Strength Indicator(RSSI) for distance estimation
in our implementation to obtain, for each node, an estimate of its 
one-hop neighborhood.  Simplifying this task was our design of the 
grid, which defined the one-hop distance between neighbors 
to be about nine meters.  We assume that the deployment error 
for each node is within a defined tolerance (see Section~\ref{sec-experiment}).

The RSSI ranging works as follows. A sensor node sends
out a radio message with a certain signal strength and one field 
of this message records the signal strength of sending.  The receiver of 
this message can measure the signal strength of the received message.
Given a model of how signal strength reduces with distance (in our
case obtained empirically), the original signal strength and received 
signal strength can be compared and the distance between the 
sender and receiver can be estimated.  The advantage of RSSI ranging 
is that it is very simple and needs no additional
hardware and little computing power. The disadvantage of RSSI ranging
is that the accuracy is poor\cite{AS04};  radio strength can be affected by the
environment, the relative angles of emplacement for a pair of sensors, 
and manufacturing variances in radio devices.  

For RSSI ranging in our 
implementation, we used a group of messages to deal with variation 
(we did not use advanced filters in our experiments, preferring to 
use the simplest of techniques).  In our experiments, we notice
that even with group size of 30, over the same distance, 
the mean RSSI reading varies considerably, as has been previously 
observed for similar hardware \cite{WC02}.  Put another way,
the mean received signal strength readings can be same over different 
distances.  To reduce the inaccuracy of RSSI ranging and get better 
one-hop neighborhood information, instead of having estimation of 
the real distance between a single pair 
of sender and receiver, we forward statistics of
the RSSI readings to the base station. 
 
Experience of other researchers \cite{WC02,PH03,ZHKS04} 
suggests that no elementary model may exist for how 
signal strength behaves as a function of distance.
Therefore, instead of trying to find an analytical model for 
RSSI, we used a machine learning approach in our investigation.
Machine learning for RSSI-based localization was previously used 
in \cite{ELM04}, where Bayesian networks were trained.

We use fuzzy membership functions \cite{D90} to 
calculate a score for each receiver by looking at average RSSI reading,
number of received messages, maximum and minimal RSSI readings, and
the rank of RSSI readings among all receivers. For each property,
we assign one fuzzy membership function. According to the classification in
\cite{D90}, we choose the membership functions
based on reliability concerns with respect to the particular problem.
From our implementation, even the very simple function forms (linear and 
triangular functions) are sufficient for this part.
Finally we decide the one-hop 
neighborhood for the sender by checking the scores of the receivers.
The score is in fact another fuzzy membership function. 
For this function, we use {\em Distance Approach}, 
one of the {\em semantic approaches} in fuzzy
membership function design. This approach concentrates more on 
practical meaning and interpretation of membership.  It is ideal 
for multi-attribute decision making \cite{LH}. This approach 
considers requirements on different attributes. It will assign 
operations for fuzzy logic. For instance, let $f_A$ and $f_B$
be fuzzy membership functions for attributes $A$ and $B$; then
fuzzy membership function for ``$A$ and $B$'' is $min(f_A,f_B)$,
and for ``very $A$'' is $f_A^2$. So we can  interpret empirical 
rules such as ``if very $A$ and $B$, then $C$ is likely'' into
functions.    

After choosing the forms of the fuzzy membership functions, we need
to decide the parameters for these functions. We conducted extensive
tests for calibration.  We performed tests with different spacing 
between nodes, relative angles, weather and terrain conditions. 
From the training data we collected from the tests, 
we determined the parameters
to best distinguish 1-hop neighbors from more distant neighbors. 
In Section~\ref{sec-analysis}, we
will show how well these functions work. 

We give an example here on how to choose parameters for
fuzzy membership functions. The fuzzy membership function $f_{avg}$ for 
an average RSSI reading $x$ to be ``numerically like" a 1-hop reading is a combination
of 3 simple linear functions. 

\[f_{avg}(x) = \left\{\begin{array}{ll}
1 & x<a\\
\frac{b-x}{b-a} & a \leq x < b \\
0 & x \geq b
\end{array}
\right. \]

In our tests, generally, if the distance is 
smaller, the reading is smaller. So a smaller reading
implies a more likely 1-hop reading.
Thus we chose the function in this form.

Here, $a$ and $b$ are parameters of this fuzzy membership function.
We need to determine these two parameters from our data. We
want to choose them to be thresholds that only a small part of 2-hop
readings are smaller than $a$ and most of the 1-hop readings are smaller
than $b$. So We chose 
the $10^{th}$ percentile of average readings over 2-hop distances for 
$a$, and the $95^{th}$ percentile of average readings over 1-hop distances
for $b$.

\begin{figure}[ht]
\centering
\includegraphics {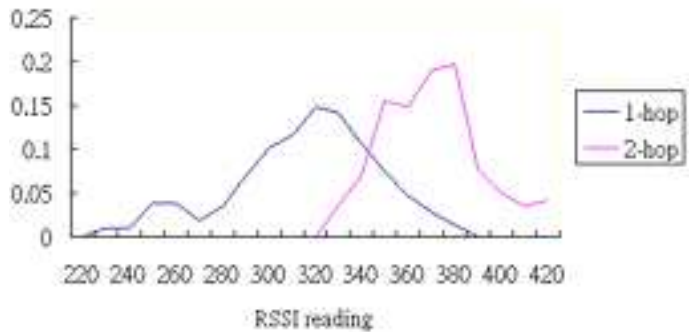}
\caption{distribution for 1-hop and 2-hop readings}
\label{distributionChart}
\end{figure}

Figure \ref{distributionChart} shows the distributions
of 1-hop and 2-hop readings. So for our training data,
$a = 343$, and $b = 361$.

An advantage of using machine learning with fuzzy membership 
function is that we do not have strong assumption for
the underlying distribution of the RSSI reading over a certain distance.

An observation from the experiment is that by 
running a $\chi^2$
goodness-of-fit test on these two distributions, we 
found that the reading distributions for a fixed
distance might not always be normal distribution,
as assumed in \cite{MLRT04,NN04}. For 2-hop readings,
it is not statistically significant to reject the
hypothesis that the distribution is normal. But for
1-hop readings, the test yields $\chi^2 = 30.11$, for
$p=0.01$, the $\chi^2$ for degree of freedom of 15 is 
30.58, so with high significance, we can reject the
hypothesis that the distribution of 1-hop readings is 
normal.

Similar to $f_{avg}$, we have fuzzy membership function
$f_{rel}$, which describes how much an average RSSI reading $x$
is ``relatively like" a 1-hop reading by checking all average RSSI readings
for the same sender. Let $s$ denote the sender of the messages, and
 $R_s$ denote the set of nodes which report
an average RSSI reading for $s$. If $|R_s|<2$,
$f_{rel}$ is always 1. When $|R_s|$ is at least 2, let $max_s$
denote the mean of the 2 strongest average RSSI readings 
reported by $R_s$. 

\[f_{rel}(x) = \left\{\begin{array}{ll}
1 &  x < 1.05max_s\\
1-\frac{x-1.05 max_s}{0.1 max_s} & 1.05max_s \leq x < 1.15 max_s \\
0 & x \geq 1.15max_s
\end{array}
\right. \]

If a receiver 
receives most of the messages in the message group
from a sender, the average RSSI reading is more reliable
than that from just a few messages. Meanwhile, from the training
data we collected, we observe that 1-hop neighbors generally receive
more messages than the 2-hop neighbors. Also with 
reduced power of radio signals and a message rate of 10 messages/s, 
we have not observed obvious message collisions, so it is not necessary
to schedule the messages.

Function $f_{num}$ is used to determine whether an 
average RSSI reading $x$ is ``like in volume" to a 1-hop reading by 
checking the number of messages on which $x$ is based.
Let $x_{num}$ denote this number, $s$ denote the sender 
of the messages, and $most_s$ denote
the most number of messages from $s$ that 
are received by a receiver.  

\[f_{num}(x) = \left\{\begin{array}{ll}
0 &  x_{num} < 0.65most_s\\
\frac{x-0.65 most_s}{0.25 most_s} & 0.65most_s \leq x_{num} < 0.9 most_s \\
1 & x_{num} \geq 0.9most_s
\end{array}
\right. \]

After calculating the fuzzy membership functions for
different attributes, we use a fuzzy rule to get a 
score for the sender/receiver pair. From the training data
we collected, we chose the rule for judging whether an average
reading of a sender/receiver pair is like a 1-hop reading to be,
it is ``numerically like", or, ``relatively like" and very ``like
in volume". So the score is calculated by the following.

\[score(x) = max(f_{avg}(x),min(f_{rel}(x),f_{num}^2(x)))\]

\subsection{Table Look-up and Refinement}

After establishing the one-hop neighborhood for every sensor node, the
algorithm constructs shortest paths for each node to all the anchors
in the segment. 
Thus it has a 4-tuple for each sensor node $q$, consisting of 
grid distances between $q$ and the anchors.  Next, a table lookup
is attempted:  if it matches a table entry of 
unoccupied position $p$, then 
the algorithm assigns sensor node $q$ to position $p$.  Due to inaccuracies
in the one-hop neighborhood determination, the lookup can fail to 
match any table entry.  To deal with lookup misses, we propose the 
following refinement: for an unoccupied position $p$, 
calculate a score for each remaining node. If a node $q$ has highest score 
among the remaining nodes and the score is above a threshold $T$, assign
$q$ to position $p$. 

Here the algorithm assigns nodes to positions instead of 
assigning positions to nodes.  It does not make much difference in one 
segment, however in a large scale network, 
the base station can receive readings of nodes from an adjacent segments. 
It then becomes likely that the base station has more nodes than positions, 
so assigning nodes to positions is more likely to yield a better result.  

Note that it is possible that two or more nodes have tuples that
match the same table entry due to inaccurate RSSI readings. 
Using our algorithm, only one will be assigned 
to that grid position depending on which one is the first
to be looked up in the table.  (Of course, RSSI ranging errors lower 
the quality of our solution, as they would any RSSI-based 
solution to localization.)

It is useful to observe that our decision to match nodes to grid 
positions, that is, to find a bijection between nodes and grid points, 
may not be appropriate for some applications.  Indeed, when two 
nodes have the same table lookup results, it can be argued that both
should receive the same grid position.  The general question of what
is a good metric for applications depending on localization quality 
is outside the scope of our research.

\subsection{Distributed Implementation within a Segment}
From our description of the algorithm, it is not hard to change the 
implementation to be fully distributed. In the centralized implementation,
the sensor nodes just need to send out a group of messages and forward
statistics of the RSSI readings to the base station. 
There is no message exchange between the sensor nodes.
After forwarding the statistics, the sensor nodes will just wait for
a grid position assignment.

In a fully distributed implementation, each sensor node will send the 
statistics of the RSSI readings to the sender of the RSSI messages. 
Prior to deployment, each node's programming includes the localization
table as a read-only constant (in current technology, there is far
more read-only memory than working RAM).
After a sender gets statistics from the receivers, it will use fuzzy 
membership functions to assign scores to each receiver and determine
which receivers are the one-hop neighbors. In stage 3, BFS spanning trees 
rooted at the anchors can be constructed using a distributed algorithm
(which could be similar to routing protocols that construct spanning trees).
The depth of a node $q$ in the tree rooted at anchor $a_i$ is the distance
between $q$ and $a_i$.  Then each node $q$ will look up the 4-tuple
in the table. If there is a match, assign the position to $q$; if not, 
assign a score to all positions and pick the position with highest score.
The difference between the centralized and distributed implementations in 
stage 3 is that distributed implementation assigns positions to nodes. So
it is possible for two nodes to think they are at the same position (as 
noted previously, this is acceptable for some applications).

In a large scale grid wireless sensor network, the network
can be heterogeneous, enabling faster communication and data processing. 
Some sensor nodes are more powerful and have larger communication range.
These nodes form the back-bone of the network,
so routing to the base station needs fewer hops.  GPS could be installed
at these nodes as well, making such nodes ideal as anchors.  If they 
have much more computing power than other sensor nodes, the 
algorithm could run on these nodes.  On the other hand, 
if there are no such powerful nodes in the network, the 
distributed implementation is more desirable.  Another advantage 
of the distributed implementation is that no multi-hop messaging 
or flooding is needed, which reduces radio traffic.

%% file: sec4.tex
\section{Experimental Result}
\label{sec-experiment}
We use the topology in Figure~\ref{segment} for the centralized
implementation. The unit length of x-axis is 4.5 meters and 
the unit length of the y-axis is 9 meters. We used 52 CrossBow MSP410
prototypes for the experiment.  These devices have some advantages
over similar sensor nodes for RSSI ranging:  the antenna is 
securely attached, centered on the device (which can reduce the 
effect of angular variance in the field strength) and tall enough
so that it does not suffer from extreme ground effects.
Fifty of these units were placed in the segment 
and two other nodes also participated
to simulate the effects of sensor nodes from an adjacent segment.
We conducted the experiment in an outdoor environment with a 
rough and grassy surface.
With the signal strength we used in the experiment, the message
can travel 15-25 meters. The deployment error of a node can be as large
as 40 cm (that is, the actual placement of the node could deviate
by as much as 40 cm from its intented target position).

\begin{figure} [ht]
\centering
\includegraphics {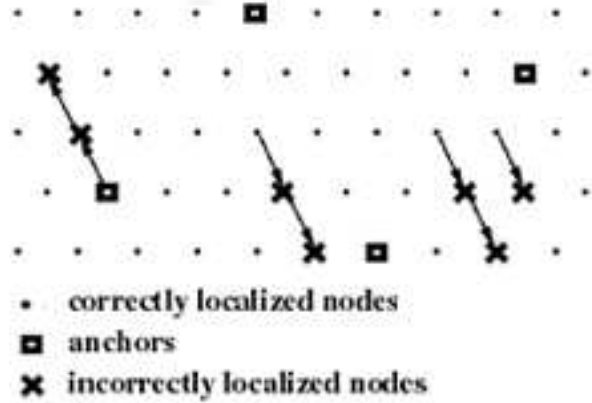}
\caption{Experimental result}
\label{result}
\end{figure}

We first tested the RSSI ranging for calibration purposes, then we 
ran the experiment to get RSSI readings. Figure~\ref{result} shows the
localization result from our implementation using
the RSSI readings. The arrows in the figure show
the real positions of the incorrectly assigned nodes. 

We propose some simple metrics here for performance evaluation purposes.
$rate_c$ is the percentage of correctly localized positions, and $avg_{error}$
is the average error of the incorrectly localized positions, 
in terms of grid distance. 
In our experiment, among the 50 positions in the
segment, we correctly assigned 43 nodes to their positions. Of all the other 7
positions, the errors are all just 1-hop. 
So in our experiment, $rate_c$= 86\% and $avg_{error}= 1$. 

In \cite{SKMKA05}, an algorithm is introduced to solve 
localization in grid wireless networks. The authors use acoustic 
Time Difference of Arrival (TDoA) technique
for distance measurement, then use a subgraph isomorphism algorithm 
and topology information to calculate the result. 
The results they show yield a $rate_c$ = 88.3\%
and $avg_{error} = 1.33$.  Thus our results are comparable to the 
acoustic method \cite{SKMKA05}, however there are advantages to using 
RSSI noted in the following paragraph.

Beyond the hardware expense of equipping nodes with sounder and microphone,
the sounder consumes power:  a buzz alone can cost energy equivalent to 
sending up to 50 messages, and numerous soundings are needed to overcome
errors in the acoustic ranging phase.
Also on the receiving end, significant signal processing is
needed to process the acoustics, which can have hardware and power cost.  
For our table lookup algorithm, the RSSI ranging needs radio 
30 messages with almost minimal signal strength.
The communication energy consumption of the two methods are similar 
since both algorithms forward the ranging data to base station. 

%% file: sec5.tex
\section{Performance Analysis}
\label{sec-analysis}
To analyze the performance, we introduce the following definitions in
for 1-hop neighborhood establishment (these definitions follow the 
classical error terminology of hypothesis testing in statistics).

\begin{definition}
If two nodes $p$ and $q$ are not 1-hop neighbors, and
they are considered by the algorithm to be 1-hop neighbors,
then the algorithm is said to make a \emph{type A error} for
$(p,q)$.
\end{definition}

\begin{definition}
If two nodes $p$ and $q$ are 1-hop neighbors, and
they are not classified by the algorithm as 1-hop neighbors,
then the algorithm is said to make a \emph{type B error} for
$(p,q)$.
\end{definition}

RSSI ranging and the fuzzy membership functions effectively have
thresholds between classifying a pair $(p,q)$ as 1-hop neighbors or 
not.  Moving this threshold one way or another can change the number
of type A and type B errors.

For our field experiment, among 121 1-hop pairs in the grid, 
we had 17 type B errors and we had 9 type A errors in 192 2-hop pairs.
Among all the 50 nodes, 24 nodes have correct 1-hop neighborhoods.
In other words, with the parameters we chose for the
fuzzy membership functions, we recognized 86.0\% of the 1-hop neighbors
and took 4.7\% of 2-hop neighbors for 1-hop neighbors.

Our experiment shows that we manage
to reduce the number of type A and type B errors using 
fuzzy membership functions
proposed in section~\ref{sec-algorithm}. This result is obtained though 
RSSI ranging data is inaccurate. We observed that even 
with 30 messages in each group,
the mean RSSI reading over the same distance could 
have a quite large variance. Distance estimation using RSSI alone 
can generate error as large as 60\%. 
In our experiment, mean RSSI reading of one group
of messages over distance of 10 meters can be 
same as the overall average reading over the distance of 16 meters. 
In fact, when we tried to 
establish the 1-hop neighborhood using RSSI distance estimation alone, 
and keep the type A error ratio at the similar level, 
we ended up having many more type B errors.
The type  A error ratio for 2-hop neighbors is 5.4\%, and ratio of
correctly recognized  1-hop neighbors
is 72.2\%.

Both type A and type B errors in the 1-hop neighborhood establishment can
affect the tuples of the nodes. But their effects are not the same.
In our topology, for most pairs of nodes, there are at least two
paths of minimum length.
If a type B error occurs and breaks one of the shortest paths, the distance
calculation is more likely to give the same result because there are other 
shortest paths. In the mean time, a type A error is more likely to change
the length of the shortest path. So we can see a type A error is more likely 
to affect the 4-tuple of a node. Thus we have the following:
\begin{quote}
{\bf Observation.}
The table lookup algorithm tolerates more type B 
errors than type A errors.
\end{quote}
For the algorithm to achieve satisfying results, 
we need as few errors as possible.
Unfortunately, there is a trade-off between the number 
of type A and type B errors.
Depending on how to choose a fuzzy membership function, the 
score assignment and the rules to pick 1-hop number, we can 
reduce the number of one type of errors, but the
number of the other type will increase. 

We also ran simulations by injecting type A and type B errors. 
It turns out that to obtain good localization result, 
we need to reduce the number of type A errors to be fewer
than 12, and number of type B errors to be fewer than 31 
in the 50 node grid.

\begin{figure}[ht]
\centering
\includegraphics {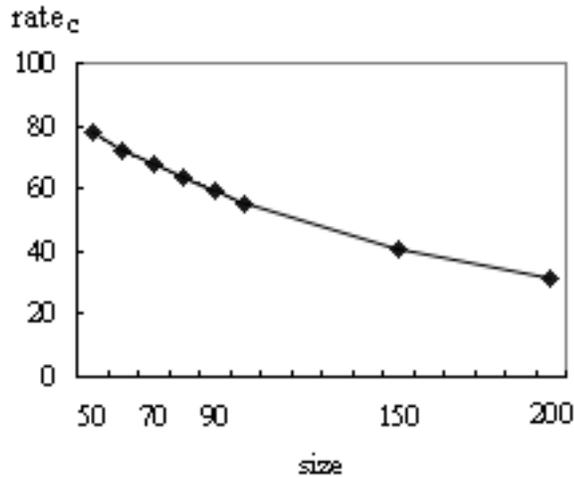}
\caption{$rate_{c}$ for different segment sizes}
\label{ratecChart}
\end{figure}

\begin{figure}[ht]
\centering
\includegraphics {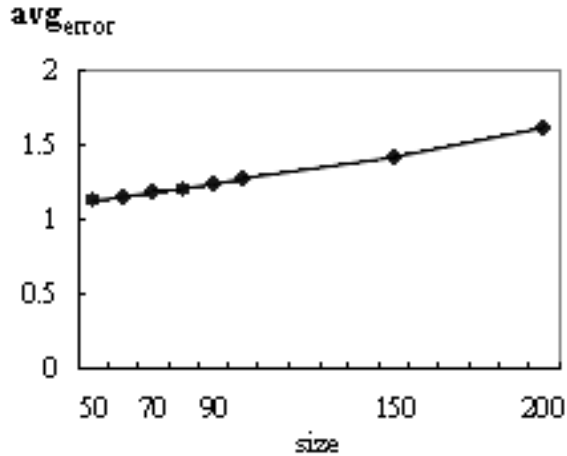}
\caption{$avg_{error}$ for different segment sizes}
\label{avgErrorChart}
\end{figure}

We explored the effect of segment size on the algorithm by simulations.
We conducted simulations for larger segments, assuming that we could get 1-hop 
neighborhood information of same accuracy as
in our experiment. We did simulations on segments with different sizes
ranging from 50 to 200(50, 60, 70, 80, 90, 100, 150, 200). All
simulations use only four anchors. For each segment 
size, we ran 1000 simulations.
Figure~\ref{ratecChart} and \ref{avgErrorChart} shows $rate_c$ and
$avg_{error}$ for different segment sizes.

With a larger segment, the accuracy of the algorithm decreases. This is 
because with longer paths, the errors in grid distance counting
will accumulate.

%% file: sec6.tex
\section{Conclusions and Future Work}
\label{conclusion}
In this paper, we confirm the advantage
of using a grid when an anchor-based RSSI 
method is used to localize a wireless sensor network.
With the grid topology information,
it is possible to have a simpler and better 
localization algorithm.
 
We introduced the table lookup localization
algorithm and implemented it with
a segment of 50 sensor nodes. The result shows
that this algorithm is efficient and energy saving.
Also by obtaining 1-hop neighborhood using aggregated
information and fuzzy membership functions, 
the result is achieved with
inaccurate RSSI ranging data. 

There is still a lot to explore. The lower bound of
1-hop neighborhood establishment is achieved by assuming
errors are mutually independent and using the current
stage 3 algorithm. We can develop a more realistic model of errors
to better simulate the RSSI readings, and improve the 
stage 3 algorithm to tolerate more errors.